\documentstyle[prl,aps,twocolumn]{revtex}

\begin{document} 
\draft 
\title{Periodic Orbit Quantization of Mixed Regular-Chaotic Systems}
\author{J\"org Main$^1$ and G\"unter Wunner$^2$}
\address{$^1$Institut f\"ur Theoretische Physik I,
         Ruhr-Universit\"at Bochum, D-44780 Bochum, Germany}
\address{$^2$Institut f\"ur Theoretische Physik und Synergetik,
         Universit\"at Stuttgart, D-70550 Stuttgart, Germany}
%
\date{\today}
\maketitle

\begin{abstract}
A general technique for the periodic orbit quantization of systems with
near-integrable to mixed regular-chaotic dynamics is introduced.
A small set of periodic orbits is sufficient for the construction of
the semiclassical recurrence function up to, in principle, infinite length.
As in our recent work the recurrence signal is inverted by means of a high 
resolution spectral analyzer (harmonic inversion) to obtain the 
semiclassical eigenenergies.
The method is demonstrated for the hydrogen atom in a magnetic field.
To our knowledge this is the first successful application of periodic
orbit quantization in the deep mixed regular-chaotic regime.
\end{abstract}


The question of how to semiclassically quantize systems with 
{\em nonintegrable} Hamiltonians had puzzled the great minds of physics in 
the early days of quantum mechanics \cite{Ein17}.
The advent of ``exact'' quantum mechanics, and, later, the availability of
more and more powerful computing resources, with the possibility of 
numerically solving, in principle, Schr\"odinger's equation for every complex 
quantum system, pushed that old question in the background for several decades.
However, the desire for deeper physical insight into ``what the computer 
understands'' \cite{Wig94} finally triggered a renaissance of semiclassical 
theory, which persists. 
It will be the objective of this Letter to contribute to solving that 
longstanding problem of semiclassical quantization of nonintegrable systems 
in the {\em mixed regular-chaotic} regime.

The breakthrough of modern semiclassical theory came when Gutzwiller proved, 
by an application of the method of stationary phase to the semiclassical
approximation of the quantum propagator, that for systems with complete 
chaotic (hyperbolic) classical dynamics the density of states can be 
expressed as an infinite sum over all (isolated) periodic orbits, thus 
laying the foundation of periodic orbit theory \cite{Gut90}.
Special methods were designed for specific systems to overcome the 
convergence problems of the semiclassical trace formula, i.e., to 
analytically continue its range of convergence to the physical domain
\cite{Cvi89,Aur92,Ber92,Bog92}.
None of these methods, however, has succeeded so far in correctly describing 
generic dynamical systems with mixed regular-chaotic phase spaces.

On the other extreme of complete integrability, it is well known that the 
semiclassical energy values can be obtained by EBK torus quantization 
\cite{Ein17}.
This requires the knowledge of all the constants of motion, which are 
not normally given in explicit form, and therefore practical EBK quantization
based on the direct or indirect numerical construction of the constants of 
motion turns out to be a formidable task \cite{Per77}.
As an alternative, EBK quantization was recast as a sum over all periodic 
orbits of a given topology on respective tori by Berry and Tabor \cite{Ber76}.
The Berry-Tabor formula circumvents the numerical construction of the 
constants of motion but usually suffers from the convergence problems of the
infinite periodic orbit sum.
 
The extension of the Berry-Tabor formula into the near-integrable (KAM) 
regime was outlined by Ozorio de Almeida \cite{Alm88} and elaborated, at 
different levels of refinement, by Tomsovic et al.\ and Ullmo et al.\ 
\cite{Tom95}.
These authors noted that in the near-integrable regime, according to the 
Poincar\'e-Birkhoff theorem, two periodic orbits survive the destruction
of a rational torus with similar actions, one stable and one hyperbolic
unstable, and worked out the ensuing modifications of the Berry-Tabor 
formula.
In this Letter we go one step further by noting that, 
with increasing perturbation, 
the stable orbit turns into an inverse hyperbolic one representing, 
together with its unstable companion with similar action, a remnant torus.
We include the contributions of these pairs of inverse hyperbolic and
hyperbolic orbits in the Berry-Tabor formula and demonstrate for a
system with mixed regular-chaotic dynamics that this procedure yields 
excellent results even in the deep mixed regular-chaotic regime.
The system we choose is the hydrogen atom in a magnetic field, which is a 
real physical system and has served extensively as a prototype for the 
investigation of ``quantum chaos'' (for reviews see \cite{Fri89,Has89,Wat93}).
To our knowledge no semiclassical quantization thus far in the mixed 
regular-chaotic region has previously been given.

The fundamental obstacle bedeviling the semiclassical quantization of systems
with mixed regular-chaotic dynamics is that the periodic orbits are neither 
sufficiently isolated, as is required for Gutzwiller's trace formula 
\cite{Gut90}, nor are they part of invariant tori, as is necessary for the 
Berry-Tabor formula \cite{Ber76}.
However, as will become clear below, it is the Berry-Tabor formula which 
lends itself in a natural way for an extension of periodic orbit quantization 
to mixed systems.
We consider systems with a scaling property, i.e., where the shape of 
periodic orbits does not depend on the scaling parameter, 
$w=\hbar_{\rm eff}^{-1}$, and the classical action $S$ scales as $S=sw$ 
with $s$ the scaled action.
For scaling systems with two degrees of freedom, which we will focus on,
the Berry-Tabor formula for the fluctuating part of the level density reads
\begin{equation}
 \varrho(w) = {1\over\pi}\, {\rm Re} \,
   \sum_{\bf M} {w^{1/2}s_{\bf M}\over M_2^{3/2}|g''_E|^{1/2}} \,
   e^{i(s_{\bf M}w-{\pi\over2}\eta_{\bf M}-{\pi\over 4})} \; ,
\label{BT_eq}
\end{equation}
with ${\bf M}=(M_1,M_2)$ pairs of integers specifying the individual periodic 
orbits on the tori (numbers of rotations per period, $M_2/M_1$ rational), and
$s_{\bf M}$ and $\eta_{\bf M}$ the scaled action and Maslov index of the 
periodic orbit ${\bf M}$.
The function $g_E$ in (\ref{BT_eq}) is obtained by inverting the Hamiltonian, 
expressed in terms of the actions $(I_1,I_2)$ of the corresponding torus,
with respect to $I_2$, viz.\ $H(I_1,I_2=g_E(I_1))=E$ \cite{Boh93}.
The calculation of $g''_E$ from the actions $(I_1,I_2)$ can be rather
laborious even for integrable and near-integrable systems, and, by definition,
becomes impossible for mixed systems in the chaotic part of the phase space.
Here we will adopt the method of Ref.\ \cite{Tom95} and calculate $g''_E$, 
for given ${\bf M}=(\mu_1,\mu_2)$, with $(\mu_1,\mu_2)$ coprime integers 
specifying the primitive periodic orbit, directly from the parameters of 
the two periodic orbits (stable (s) and hyperbolic unstable (h)) that 
survive the destruction of the rational torus $\bf M$, viz.
\begin{equation}
 g''_E = {2\over\pi \mu_2^3\Delta s}
   \left({1\over\sqrt{\det(M_{\rm s}-I)}}
       + {1\over\sqrt{-\det(M_{\rm h}-I)}}\right)^{-2} \, ,
\label{g2_eq}
\end{equation}
with
\begin{equation}
 \Delta s = {1\over 2} (s_{\rm h} - s_{\rm s})
\label{Ds_eq}
\end{equation}
the difference of the scaled actions, and $M_{\rm s}$ and $M_{\rm h}$ the 
monodromy matrices of the two orbits.
The action $s_{\bf M}$ in (\ref{BT_eq}) is to replaced with the mean action
\begin{equation}
 \bar s = {1\over 2} (s_{\rm h} + s_{\rm s}) \; .
\label{s_mean_eq}
\end{equation}
Eq.\ \ref{g2_eq} is an approximation which becomes exact in the limit
of an integrable system.

It is a characteristic feature of systems with mixed regular-chaotic dynamics
that with increasing nonintegrability the stable orbits turn into inverse 
hyperbolic unstable orbits in the chaotic part of the phase space.
These orbits, although embedded in the fully chaotic part of phase space,
are remnants of broken tori.
It is therefore natural to assume that Eqs.\ \ref{BT_eq} and \ref{g2_eq} can 
even be applied when these pairs of inverse hyperbolic and hyperbolic orbits 
are taken into account, i.e., more deeply in the mixed regular-chaotic regime.

It should be noted that the difference $\Delta s$ between the actions of the 
two orbits is normally still small, and it is therefore more appropriate 
to start from the Berry-Tabor formula for semiclassical quantization in that 
regime than from Gutzwiller's trace formula, which assumes well-isolated 
periodic orbits.
It is also important to note that the Berry-Tabor formula does not require
an extensive numerical periodic orbit search.
The periodic orbit parameters $s/M_2$ and $g''_E$ are smooth functions
of the rotation number $M_2/M_1$, and can be obtained for arbitrary periodic 
orbits with coprime integers $(M_1,M_2)$ by interpolation between ``simple''
rational numbers $M_2/M_1$.

\begin{figure}[t]
\vspace{9.1cm}
\includegraphics{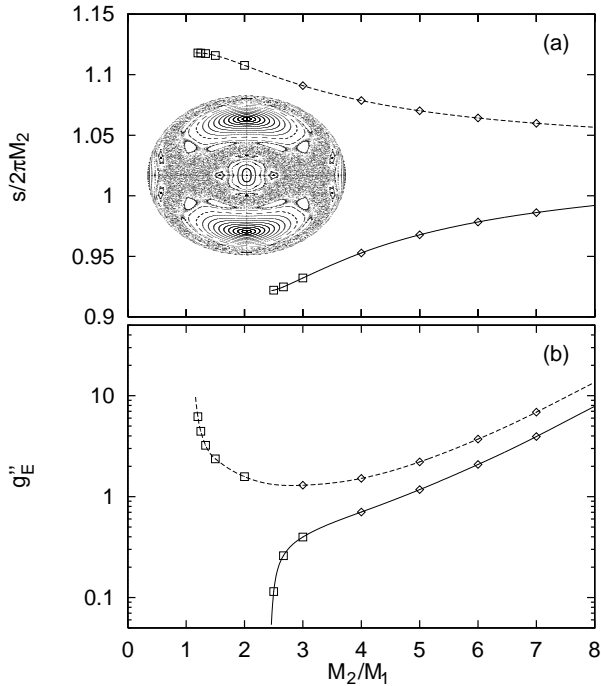}
\caption{\label{fig1} 
(a) Action $s/2\pi M_2$ and (b) second derivative $g''_E$ as a function of 
the frequency ratio $M_2/M_1$ for the rotator (solid lines) and vibrator 
(dashed lines) type motion of the hydrogen atom in a magnetic field at 
scaled energy $\tilde E=-0.4$.
Inset: Poincar\'e surface of section in semiparabolical coordinates
$(\mu,p_\mu;\nu=0)$.
}
\end{figure}

We now demonstrate the high quality of the extension of Eqs.\ \ref{BT_eq} 
and \ref{g2_eq} to pairs of inverse hyperbolic and hyperbolic periodic orbits 
for a physical system that undergoes a transition from regularity to chaos, 
namely the hydrogen atom in a magnetic field.
This is a scaling system, with $w=\gamma^{-1/3}=\hbar_{\rm eff}^{-1}$ the 
scaling parameter and $\gamma=B/(2.35\times 10^5\, {\rm T})$ the magnetic 
field strength in atomic units.
Introducing scaled coordinates $\gamma^{2/3}{\bf r}$ and momenta 
$\gamma^{-1/3}{\bf p}$ and choosing the projection of the angular momentum 
on the magnetic field axis $L_z=0$ one arrives at the scaled Hamiltonian 
\begin{equation}
   \tilde H = {1\over 2}{\bf p}^2 - {1\over r} + {1\over 8}(x^2+y^2)
 = \tilde E \; ,
\end{equation}
with $\tilde E=E\gamma^{-2/3}$ the scaled energy.
At low energies $\tilde E<-0.6$ a Poincar\'e surface of section analysis 
of the classical dynamics \cite{Has89} exhibits two different torus 
structures related to a ``rotator'' and ``vibrator'' type motion.
The separatrix between these tori is destroyed at a scaled energy of
$\tilde E \approx -0.6$, and the chaotic region around the separatrix grows
with increasing energy.
At $\tilde E=-0.127$ the classical phase space becomes completely chaotic.
We investigate the system at scaled energy $\tilde E=-0.4$, where about
40\% of the classical phase space volume is chaotic 
(see inset in Fig.\ \ref{fig1}), i.e.\ well in the region of mixed dynamics.
We use 8 pairs of periodic orbits to describe the rotator type motion in 
both the regular and chaotic region.
The results for the periodic orbit parameters $s/2\pi M_2$ and $g''_E$ are
presented as solid lines in Fig.\ \ref{fig1}.
The squares on the solid lines mark parameters obtained by pairs of stable
and unstable periodic orbits in the regular region of the phase space.
The diamonds mark parameters obtained by pairs of two unstable (inverse
hyperbolic and hyperbolic) periodic orbits in the chaotic region of phase 
space.
The cutoff is related to the winding angle $\phi=1.278$ of the fixed point of 
the rotator type motion, i.e., the orbit perpendicular to the magnetic field 
axis, $(M_2/M_1)_{\rm cutoff}=\pi/\phi=2.458$.
The solid lines have been obtained by spline interpolation of the data points.
In the same way the periodic orbit parameters for the vibrator type motion
have been obtained from 11 pairs of periodic orbits (see the dashed lines
in Fig.\ \ref{fig1}).
The cutoff at $M_2/M_1=\pi/\phi=1.158$ is related to the winding 
angle $\phi=2.714$ of the fixed point of the vibrator type motion, i.e.,
the orbit parallel to the field axis.

With the data of Fig.\ \ref{fig1} we have all the ingredients at hand to 
calculate the semiclassical density of states $\varrho(w)$ in Eq.\ \ref{BT_eq}.
The periodic orbit sum includes for both the rotator and vibrator type motion
the orbits with $M_2/M_1>(M_2/M_1)_{\rm cutoff}$.
For each orbit the action and 
\newpage
\phantom{}
\noindent
the function $g''_E$ is obtained from the 
spline interpolations.
The Maslov indices are $\eta_{\bf M}=4M_2-M_1$ for the rotator and
$\eta_{\bf M}=4M_2+2M_1-1$ for the vibrator type orbits.
However, the problem is to extract the semiclassical eigenenergies from
Eq.\ \ref{BT_eq} because the periodic orbit sum does not converge.
We therefore adopt the method of Refs.\ \cite{Mai97,Mai98} where we proposed
to adjust the semiclassical recurrence signal, i.e., the Fourier transform
of the weighted density of states $w^{-1/2}\varrho(w)$
(Eq.\ \ref{BT_eq})
\begin{equation}
 C^{\rm sc}(s) = \sum_{\bf M} {\cal A}_{\bf M} \delta(s-s_{\bf M}) \; ,
\label{C_sc}
\end{equation}
with the amplitudes being determined exclusively by periodic orbit quantities,
\begin{equation}
 {\cal A}_{\bf M} = {s_{\bf M}\over M_2^{3/2}|g''_E|^{1/2}} \, 
   e^{-i{\pi\over2}\eta_{\bf M}} \; ,
\end{equation}
to the functional form of its quantum mechanical analogue
\begin{equation}
 C^{\rm qm}(s) = -i \sum_k d_k e^{-iw_k s} \; ,
\label{C_qm}
\end{equation}
where the $w_k$ are the quantum eigenvalues of the scaling parameter, 
and the $d_k$ are the multiplicities of the 
\newpage
\phantom{}
\begin{figure}[t]
\vspace{9.1cm}
\includegraphics{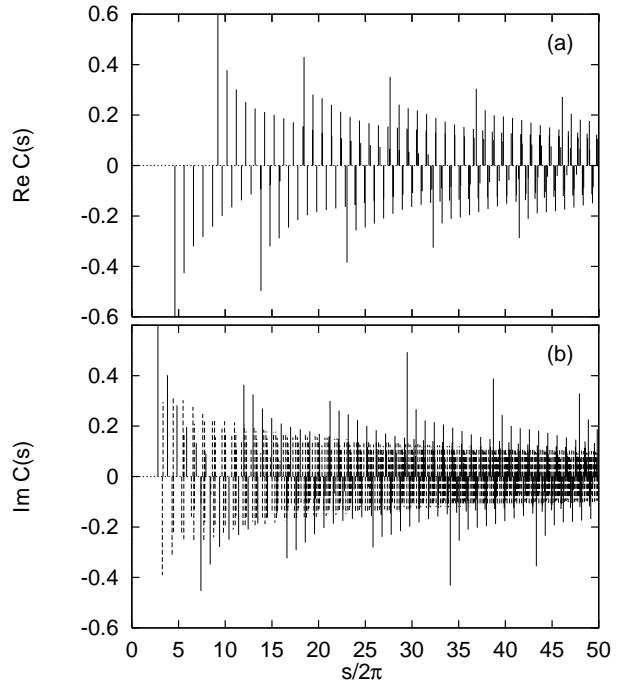}
\caption{\label{fig2} 
Semiclassical recurrence signal $C^{\rm sc}(s)$ for the hydrogen atom in a 
magnetic field at scaled energy $\tilde E=-0.4$.
Solid and dashed sticks: Signal from the rotator and vibrator type motion,
respectively.
}
\end{figure}
\noindent
eigenvalues ($d_k=1$ for nondegenerate states).
The frequencies obtained from this procedure are interpreted as the 
semiclassical eigenvalues $w_k$.
The technique used to adjust (\ref{C_sc}) to (\ref{C_qm}) is harmonic 
inversion \cite{Mai98}.

For the hydrogen atom in a magnetic field part of the semiclassical 
recurrence signal $C^{\rm sc}(s)$ at scaled energy $\tilde E=-0.4$ is 
presented in Fig.\ \ref{fig2}.
The solid and dashed peaks mark the recurrencies of the rotator and vibrator
type orbits, respectively.
Note that $C^{\rm sc}(s)$ can be easily calculated even for long periods $s$
with the help of the spline interpolation functions in Fig.\ \ref{fig1}.
By contrast, the construction of the recurrence signal for Gutzwiller's 
trace formula usually requires an exponentially increasing effort for the
numerical periodic orbit search with growing period.

\begin{figure}[t]
\vspace{9.3cm}
\includegraphics{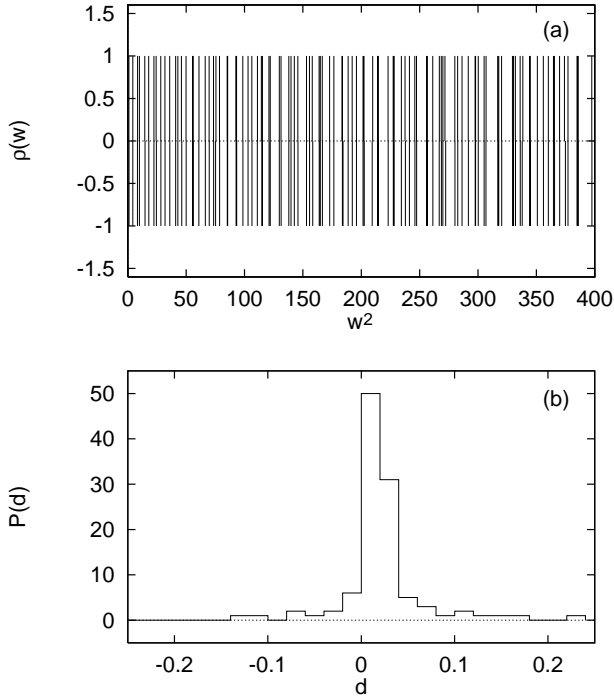}
\caption{\label{fig3} 
(a) Semiclassical and quantum mechanical spectrum of the hydrogen atom 
in a magnetic field at scaled energy $\tilde E=-0.4$.
(b) Distribution $P(d)$ of the semiclassical error in units of the mean 
level spacing,
$d=(w_{\rm qm}-w_{\rm sc})/\Delta w_{\rm av}$ for the lowest 106 eigenvalues.
}
\end{figure}

We have analyzed $C^{\rm sc}(s)$ by the harmonic inversion technique 
in the region $0<s/2\pi<200$.
The resulting semiclassical spectrum of the lowest 106 states with
eigenvalues $w<20$ is shown in the upper part of Fig.\ \ref{fig3}a.
For graphical purposes the spectrum is presented as a function of the squared
scaling parameter $w^2$, which is equivalent to unfolding the spectrum
to constant mean level spacing.
For comparison the lower part of Fig.\ \ref{fig3}a shows the exact quantum
spectrum.
The semiclassical and quantum spectrum are seen to be in excellent agreement,
and deviations are less than the stick widths for nearly all states.
The distribution $P(d)$ of the semiclassical error with 
$d=(w_{\rm qm}-w_{\rm sc})/\Delta w_{\rm av}$ the error in units
\newpage
\phantom{}
\noindent
of the mean level spacing, $\Delta w_{\rm av}=1.937/w$, is presented in 
Fig.\ \ref{fig3}b.
For most levels the semiclassical error is less than 4\% of the mean level
spacing, which is typical for a system with two degrees of freedom 
\cite{Boa94}.

The accuracy of the results presented in Fig.\ \ref{fig3} seems to be
surprising for two reasons.
First, we have not exploited the mean staircase function $\bar N(w)$,
i.e., the number of eigenvalues $w_k$ with $w_k<w$, which is a basic
requirement of some other semiclassical quantization techniques for bound 
chaotic systems \cite{Aur92,Ber92}.
Second, as mentioned before, Eq.\ \ref{g2_eq} has been derived for 
near-integrable systems, and is only an approximation, in particular, for 
mixed systems.
We have not taken into account any more refined extensions of the 
Berry-Tabor formula (\ref{BT_eq}) as discussed, e.g., in Ref.\ \cite{Tom95}.
The answer to the second point is that the splitting of scaled actions of 
the periodic orbit pairs used in Fig.\ \ref{fig1} does not exceed 
$\Delta s=0.022$, and therefore for states with $w<20$ the phase shift 
between the two periodic orbit contributions is $w\Delta s=0.44$, at most.
For small phase shifts the extension of the Berry-Tabor formula to 
near-integrable systems results in a damping of the {\em amplitudes} of the 
periodic orbit recurrence signal in Fig.\ \ref{fig2} but seems not to effect 
the {\em frequencies}, i.e., the semiclassical eigenvalues $w_k$ obtained by 
the harmonic inversion of the function $C^{\rm sc}(s)$.

In conclusion, we have presented a solution to the fundamental problem of 
semiclassical quantization of nonintegrable systems in the mixed 
regular-chaotic regime.
We have demonstrated the excellent quality of our procedure for the hydrogen 
atom in a magnetic field at a scaled energy $\tilde E=-0.4$, where about 
40\% of the phase space volume is chaotic.
The lowest 106 semiclassical and quantum eigenenergies have been shown
to agree within a few percent of the mean level spacings.
Obviously, it will be straightforward, and rewarding, to apply the method 
to other systems with mixed dynamics.

\smallskip
This work was supported in part by the Sonder\-forschungs\-be\-reich 
No.\ 237 of the Deutsche For\-schungs\-ge\-mein\-schaft.
J.M.\ thanks the Deutsche For\-schungs\-ge\-mein\-schaft for a 
Habilitandenstipendium (Grant No.\ Ma 1639/3).

\end{document}